\documentclass[10pt]{article}
\usepackage[letterpaper]{geometry}
\usepackage{hicss}
\usepackage{times}
\usepackage[none]{hyphenat}
\usepackage{url}
\usepackage{latexsym}
\usepackage{graphicx}
\usepackage{subcaption}
\usepackage{caption}
\graphicspath{{images/}}
\usepackage[
    style=ieee,
    maxnames=1,    
    minnames=1,    
    maxbibnames=1 
]{biblatex}
\usepackage{enumitem}
\usepackage{svg}

\usepackage{indentfirst}
\usepackage{xcolor}
\usepackage{amsmath}
\usepackage{graphicx}
\usepackage{float}
\usepackage{algorithmic}
\usepackage[linesnumbered,ruled]{algorithm2e}
\usepackage{array}
\usepackage{tabularx}
\usepackage{optidef} 
 
\usepackage{booktabs}
\usepackage{amssymb}
\usepackage{multirow}
\usepackage{makecell}
\usepackage{threeparttable}
\usepackage{indentfirst}
\usepackage{hyperref}
\usepackage{enumitem,kantlipsum}

\usepackage{xcolor}
\usepackage{hyperref}
\usepackage{multicol}
\usepackage{fancyhdr}

\usepackage[subtle]{savetrees}

\newcolumntype{Y}{>{\centering\arraybackslash}X}

\usepackage{mathabx}

\usepackage{caption}
\usepackage{subcaption}
\title{\textbf{Aggregate Modeling of Air-Conditioner Loads \\ Under Packet-based Control with Both On and Off Grid Access Requests}\thanks{This material is based upon work supported by the U.S. Department of Energy under Award Number DE-CR0000039. Any opinions, findings, conclusions, or recommendations expressed herein are those of the authors and do not necessarily reflect the views of the U.S. Department of Energy or the United States Government.}}
\author{
    \begin{minipage}{0.45\textwidth}
        \centering
        Mohammad Hassan \\
        University of Vermont \\
        {mhassan3@uvm.edu}
    \end{minipage}%
    \hspace{0.1\textwidth} 
    \begin{minipage}{0.45\textwidth}
        \centering
        Mads R. Almassalkhi \\
        University of Vermont \\
        {malmassa@uvm.edu}
    \end{minipage}
    \vspace{-12mm}
}
\date{}
\vspace{-8mm}
\begin{document}
\maketitle
\thispagestyle{fancy} 
\begin{abstract}
\vspace{-1.0em}
\noindent Coordination of distributed energy resources (DERs) can engender flexibility necessary to improve grid reliability. Packetized Energy Management (PEM) is a method for coordinating DERs, such as thermostatically controlled loads (TCLs) and electric vehicles, within customer quality-of-service (QoS) limits. In PEM, a DER uses local information to offer flexibility by sending a request to the DER coordinator to turn-ON or turn-OFF. Much work has focused on modeling and analyzing aggregations of DERs under PEM with fixed packet durations and only turn-ON requests. Different recent efforts to enable variable packet lengths have shown an increase in available flexibility and ramping capability, but have not been modeled in aggregate, which limits systematic analyses. To address this issue, this paper presents a new aggregate bin-based (macro) model of PEM loads that incorporates both turn-ON and turn-OFF request features, enabling the model to accurately characterize the capability of the fleet of DERs to track a power reference signal, population temperature dynamics, aggregate request rates, and variable packet lengths. Simulation-based validation is performed against an agent-based (micro) model to evaluate robustness and quantify model accuracy. Finally, the distribution of variable packet lengths from macro-model simulations are applied to inform past work on PEM with randomized packet lengths. 
\end{abstract}
\vspace{2pt}
\vspace{-0.7em}
\section{Introduction}
\vspace{-1.0em}
\noindent 
Balancing power supply and demand is key to grid reliability, and increased renewable generation has made this more complex. In response, efforts have focused on improving system reliability via either traditional bulk power investments (e.g., power generation and transmission) or maximizing utilization of existing infrastructure by leveraging flexible demand-side resources to actively enhance stability.

Central to the load-based approaches is \emph{Demand Response}—defined as the \emph{temporary, voluntary, and active} participation of loads to modify consumption patterns, hence contributing to a more resilient power grid \cite{10938867}. This paradigm shift recognizes loads not merely as passive consumers but as dynamic assets capable of providing ancillary services ~\cite{callaway2010achieving}.
The inherent thermal storage capacity of residential TCLs particularly HVAC systems and electric water heaters provides a natural physical basis for their effective participation in demand response initiatives. The property of thermal inertia allows for temporary and instantaneous power consumption adjustments while maintaining service quality within acceptable thermal comfort boundaries.

Although individual TCLs at a residential level possess relatively low power ratings, their coordinated aggregate control can provide substantial contributions to grid resilience.
Several control strategies~\cite{pem_conv,somya,8810635} have been developed to orchestrate control of TCL populations, balancing resilience objectives with end-user comfort obligations. These approaches demonstrate that properly coordinated TCL fleets can maintain customer satisfaction while delivering valuable grid services.

Analyzing the system-level impact, tractability, and performance of thousands of TCLs is challenging when approached individually. Hence, a population-level model of their aggregate behavior can facilitate the development of effective control laws and strategies ~\cite{Badshah}. 
Early aggregate modeling approaches for TCLs include the work presented in ~\cite{fokker_plank}, where Fokker–Planck-based stochastic differential equations were employed to characterize the aggregate system dynamics. Subsequently, authors in ~\cite{Badshah} developed a partial differential equation (PDE) framework that models device populations by discretizing the temperature space, thereby enabling representation of the population’s evolution in the form of bins. In an alternative approach, work presented in ~\cite{kochs} formulated a probabilistic aggregate model that also leverages temperature quantization to capture the evolution of large populations of devices using the concept of Markov chains. 

This work builds upon established efforts in PEM fleet modeling to develop a more complete and accurate population model. The first agent-level implementation of PEM, henceforth referred to as the \textit{conventional} PEM model, was introduced in \cite{pem_conv}. In this context, "agent-level" refers to the individual modeling and simulation of devices. The work in~\cite{leke} extended the conventional model by allowing TCLs to issue turn-OFF requests in addition to the turn-ON requests, improving tracking performance at the device level, though without developing a corresponding population-level model. However,~\cite{leke} implicitly introduces the concept of variable packet lengths, which was also discussed in~\cite{braham}, although without a formal theoretical foundation. In an effort to develop an aggregate model for PEM, ~\cite{main_macro} proposed a model—henceforth referred to as the \textit{conventional macro model} for managing fleets of DERs under PEM. However, this formulation did not incorporate device turn-OFF capabilities. Subsequent literature on macro modeling for PEM has largely built upon the foundation established in~\cite{main_macro}.
Prior studies have analyzed key system characteristics, such as nominal power consumption and the observability of DER fleets managed under PEM. Other research has extended the macro model to account for heterogeneous DER fleets in reference-tracking scenarios. Additionally, the development of Virtual Battery models tailored to PEM has been proposed. This work further extends the aggregate modeling efforts for PEM, and as such, the key contributions of this paper are as follows:
\begin{enumerate}[itemsep=1pt, parsep=1pt, topsep=2pt, partopsep=0pt]
    \item We extend the conventional macro model of PEM to develop an \textit{enhanced aggregate (macro)} model to additionally capture the effects of turn-OFF requests from the PEM-enabled devices at an aggregate level. 
    \item The new enhanced macro model systematically addresses the concept of variable packet length to inform design strategies specifically within conventional PEM framework to achieve improved performance without the use of OFF-requests.
    \item Simulation-based analysis is conducted to characterize model performance, accuracy and robustness.
\end{enumerate}

\section{Preliminaries}
\vspace{-1.0em}
\noindent In this section, we review the fundamentals of PEM and the basics of first-order mathematical modeling for TCLs, with a specific focus on air conditioning (AC) units. While the equations are generalizable to TCLs, they are presented here in the context of cooling thermostats operating on a hot day. 

\subsection{PEM fundamentals}  
Packetized Energy Management (PEM) is a decentralized, device-driven coordination strategy for managing DERs, where devices interact with the grid by requesting fixed-duration energy packets, each spanning a defined time interval or also referred to as \textit{packet length}. Access requests are made \textit{asynchronously}, \textit{probabilistically}, and \textit{anonymously}, with the request probability determined by the local state of each device and its need for energy. The acceptance or rejection of these requests is governed by a \textit{system-level objective}, such as shaping the power of the DER fleet to track a desired reference power signal. As illustrated in Fig. 1, the PEM coordinator accepts or rejects device requests based on the error between the aggregate power demand and the reference signal provided by a higher-level market coordinator. Communication overhead remains limited, as devices only send access requests, and the anonymity of these requests ensures fair and equitable access to the grid.


For a device with setpoint temperature $T_{\text{set}}$, CQoS is maintained by keeping the 
temperature within the deadband $D$, i.e., the interval 
$[T_{\text{set}} - \tfrac{D}{2},\, T_{\text{set}} + \tfrac{D}{2}]$.PEM maintains CQoS by sending devices outside this range into an opt-out mode, granting temporary unconditional grid access until device temperature returns into the deadband range. 

Given an air-conditioning (AC) unit \( n \), its indoor air temperature at time $t$, \( T_n(t) \), evolves dynamically according to a first-order equivalent thermal parameter (ETP) model.
\vspace{-0.5em}  
\begin{multline}
\dot{T}_n(t) = \frac{1}{R^{n}_{\text{eq}} C^{n}_{\text{eq}}} \big( T_{\text{amb}}(t) - T_n(t) \\
- \eta P_{\text{rate}} m_n(t) R^{n}_{\text{eq}} \big) + w(t),
\label{thermalmodel}
\end{multline}
where $T_{\text{amb}}$ is the ambient temperature, and $R_{\text{eq}}^n$, $C_{\text{eq}}^n$, $\eta$, and $P_{\text{rate}}$ are the unit's thermal resistance, capacitance, coefficient of performance, and rated power, respectively. Device parameters varying across the population for this study are indexed by $n$ in \eqref{thermalmodel}(see Section 4).

The operating state of the AC unit is $m_n(t)\in\{0,1\}$ with $m_n(t)=1$ ($m_n(t)=0$), if the AC is ON (OFF) at time $t$. \( w(t) \) represents a stochastic disturbance process that captures random perturbations to $T_n(t)$. The time-constant of the system is $\tau_{n} = R^{n}_{\text{eq}} C^{n}_{\text{eq}}$ and, after applying discretization with timestep $\Delta t$, the  difference equation becomes
\begin{align}
T_n[k+1] &= \left(1 - \frac{\Delta t}{\tau_{n}}\right) T_n[k] \nonumber \\
&\quad + \frac{\Delta t}{\tau_{n}} \left( T_{\text{amb}}[k] - \eta P_{\text{rate}} R^{n}_{\text{eq}} m_n[k] \right) \nonumber \\
&\quad + \Delta t \, w[k].
\end{align}
The conventional control mode of the HVAC units, sometimes also called the \textit{hysteresis} control, decides $m$ based on the following logic:
\begin{equation}
\begin{aligned}
m[k+1] = 
\begin{cases} 
1 & \text{if } m[k] = 0 \text{ and } T_{n}[k] \geq T_{\text{max}}, \\
0 & \text{if } m[k] = 1 \text{ and } T_n[k] \leq T_{\text{min}}, \\
m[k] & \text{otherwise}.
\end{cases}
\end{aligned}
\end{equation}
\begin{figure}
    \centering
    \includegraphics[width=1\columnwidth]{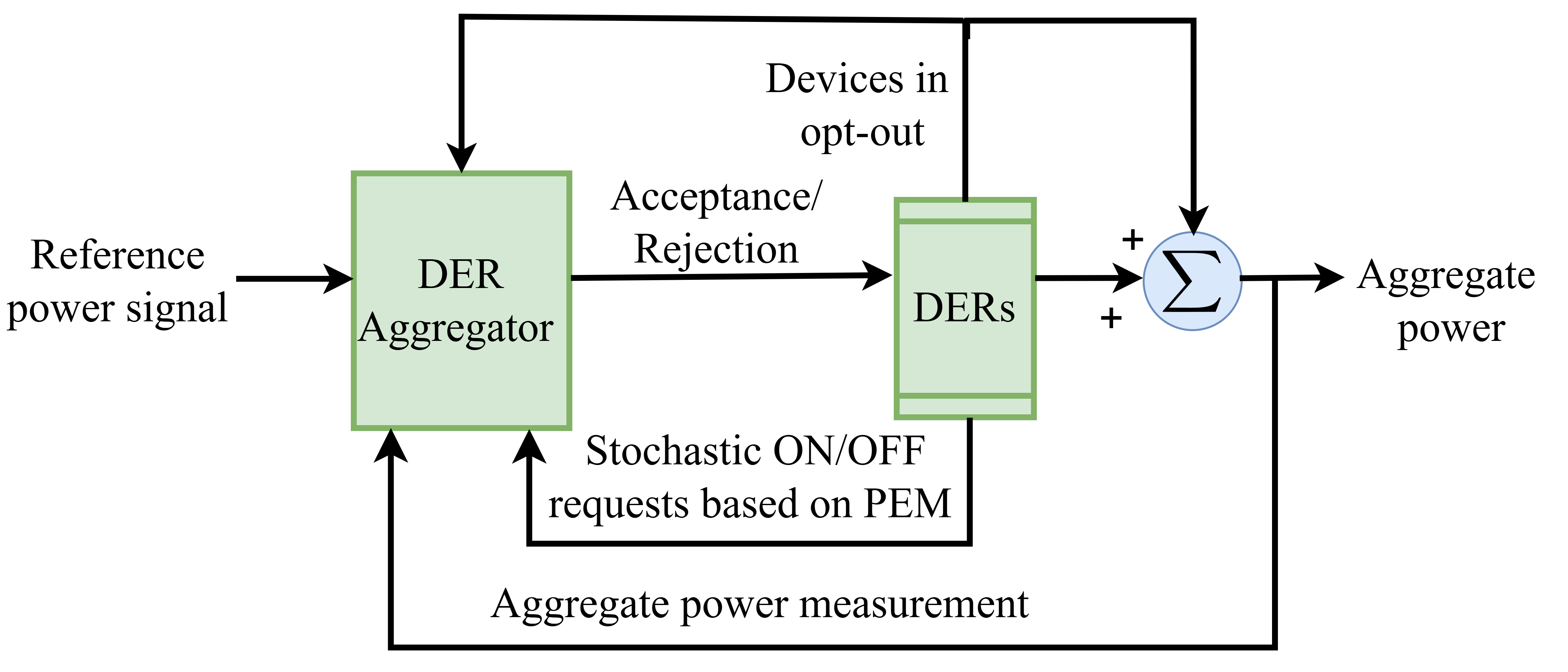}
    \caption{Typical architectural structure and components involved in PEM-based coordination of DERs}
    \label{fig:enter-label}
\end{figure}
Hysteresis logic enforces device state changes at the upper and lower bounds of the deadband to maintain thermal comfort. Within the deadband region, thermal inertia enables flexibility in manipulating the switching signal $m(t)$ to control aggregate power. Within the deadband, PEM controls OFF-ON  transitions by requiring devices to submit access requests to the controller to turn-ON or OFF instead of switching autonomously. As such, PEM employs a probabilistic framework to generate requests that reflect the need for energy; for example, higher indoor air temperature corresponds to a higher likelihood of submitting an energy request for an AC unit.

The probability of a request being made  by an AC unit $n$  in given time interval $\Delta t$, with the local, sensed air temperature \( T_{n}[k] \), is given by the cumulative distribution function(CDF) as 
\begin{equation}
    P_{\text{req-on}}(T_n[k]) = 1 - e^{-\mu _{\text{on}}(T_n[k]) \Delta t},
    \label{p_turn_on}
\end{equation}
as illustrated in Fig. 2(a). The value of $\mu(T_{n}[k])$ depends on temperature and is defined as, 

\begin{equation}
\mu_{on}(T_n[k]) = 
\begin{cases} 
0, & \text{if } T_n[k] \leq T_{\text{min}}, \\
\frac{T_n[k] - T_{\text{min}}}{T_{\text{max}} - T_n[k]} \, m_{\text{R-on}}, & \text{if } T_n[k] \in (T_{\text{min}}, T_{\text{max}}), \\
\infty, & \text{if } T_n[k] \geq T_{\text{max}},
\end{cases}
\label{eq:turn-on}
\end{equation}
where the ON request rate, \( m_{\text{R-on}} \), is inversely related to the mean time to request, (\( \text{MTTR}^{\text{on}} \)), which represents the delay before a device can attempt to send another ON-request after a previous one has been denied.
Measured in units of Hz, \( m_{\textit{R-on}} \) has implications on the bandwidth requirements of the communication network.

Authors in \cite{leke} proposed an enhancement to the conventional PEM controller ~\cite{pem_conv} by allowing DERs that are currently ON and consuming energy packets to submit OFF-requests as well. This extension introduces a time-based mechanism, enabling ON state TCLs to request early termination of their packets. 
Specifically, the probability that a TCL $n$ which has consumed \( \delta - t_n[k] \) of its total packet length \( \delta \) makes an OFF-request over time interval $\Delta t$ is illustrated in Fig. 2(b) and is defined as
\begin{equation}
 P_{\text{req-off}}(t_{n}[k]) = 1 - e^{\gamma(t_{n}[k]) \Delta t}   
 \label{p_req_off}
\end{equation}
where \( \gamma(t_{n}[k]) \) is defined as:
\begin{equation}
\gamma(t_{n}[k]) = 
\begin{cases}
0, & \text{if } t_n[k] \leq t_{\text{on}}^{\text{min}} \\
t_n[k], & \text{if } t_{\text{on}}^{\text{min}} < t_n[k] \leq t_{\text{on}}^{\text{max}} \\
t_{\text{on}}^{\text{max}}, & \text{if } t_n[k] > t_{\text{on}}^{\text{max}}.
\end{cases}
\end{equation}
This enhancement incorporates compressor lockout constraints by enforcing \( t_{\text{on}}^{\text{min}} \) as the minimum time a device must remain ON before requesting to turn OFF, and \( t_{\text{on}}^{\text{max}} \) as the maximum duration it can remain ON.
It was observed that inclusion of OFF-requests enhances DER fleet flexibility and, thus, the tracking performance offered by the fleet.

\begin{figure}[t]
    \centering
    \begin{subfigure}[t]{0.49\linewidth}
        \centering
        \includegraphics[width=\linewidth]{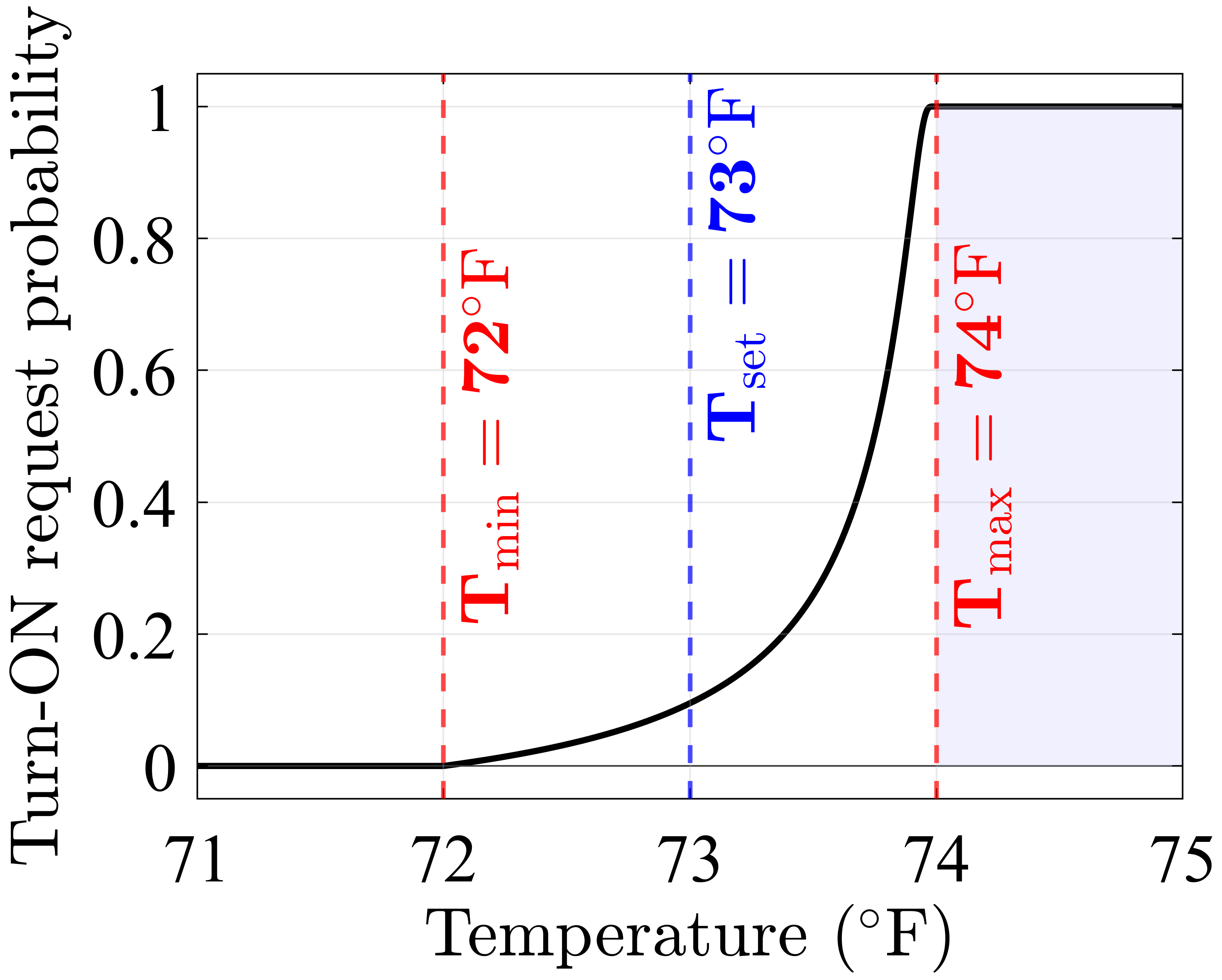}
        \caption{}
        \label{fig:subfig-turnon}
    \end{subfigure}
    \hfill
    \begin{subfigure}[t]{0.49\linewidth}
        \centering
        \includegraphics[width=\linewidth]{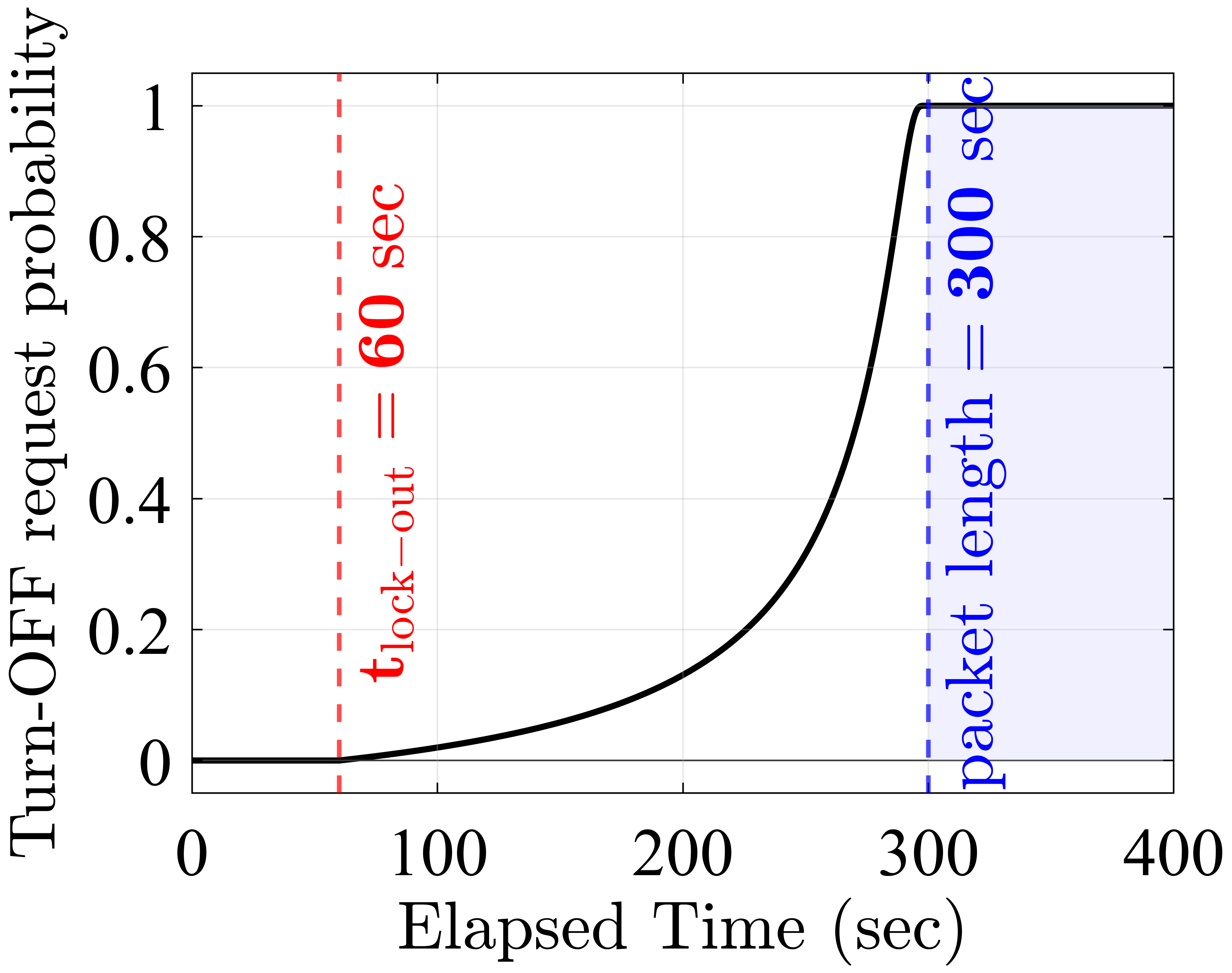}
        \caption{}
        \label{fig:subfig-turnoff}
    \end{subfigure}
    \caption{Exponential CDFs used for probabilistic ON and OFF- request decisions. (a) Temperature-based exponential CDF for turn-ON requests. (b) Time-based exponential CDF for turn-OFF requests.}
    \label{fig:exp-cdfs}
\end{figure}
\section{Aggregate modeling}
\vspace{-0.5em}
\noindent The aggregate modeling of a TCL fleet involves discretizing the continuous operating temperature range (or deadband) into \( N \) bins or states. Since each TCL can be either ON or OFF, we define an additional \( N \) bins over the same temperature range, resulting in a total of \( 2N \) discrete states. The macro-model is thus defined over a finite state space \( \mathcal{X} = \{x_1, \dots, x_{2N}\} \), where each \( x_j \) corresponds to a temperature bin associated with either an ON or an OFF-device.
We define a sequence of random variables \( \{X_k\}_{k \geq 0} \), with \( X_k : \Omega \rightarrow \mathcal{X} \), representing the state of a randomly selected device at time \( k \), where \( \Omega \) denotes the underlying sample space of continuous temperature values. For each state \( x_j \in \mathcal{X} \), we define the probability \( q_j[k] = \mathbb{P}(X_k = x_j) \), where \( j \in \{1, \dots, 2N\} \). The vector \( q[k] := \left(q_{1,\text{on}}[k], \dots, q_{N,\text{on}}[k],\ q_{1,\text{off}}[k], \dots, q_{N,\text{off}}[k]\right)^\top \in \mathbb{R}^{2N} \) represents the probability mass function of \( X_k \). For a sufficiently large fleet of DERs, each \( q_j[k] \) approximates the fraction of devices in state \( x_j \) at time \( k \), thereby allowing the macro-model to capture the aggregate fleet behavior.

The conventional hysteresis control of the TCL fleet represented as a Markov chain is given by
\begin{equation}
q[k+1] = Mq[k].
\label{eq:markov}
\end{equation}
The matrix \( M \in \mathbb{R}^{2N \times 2N} \) represents the state probability transition matrix, where each element \( P_{ij} \), with \( i, j \in \{1, \dots, 2N\} \), denotes the probability of transitioning from state \( j \) to state \( i \). Assuming a small sampling interval and sufficiently large bin width, \( M \) typically exhibits a structurally sparse form, with nonzero entries primarily along its tri-diagonal positions. The transition matrix can be constructed either analytically or through agent-based simulations of the DER fleet over a sufficiently long duration. For a comprehensive discussion on the derivation and structure of this matrix, reader is referred to \cite{kochs, NazirHiskens17_NoiseParameterHeterogeneityInAggregateModelsOfThermostatically}.

The PEM aggregate model extends the Markov chain hysteresis model in~\eqref{eq:markov} by introducing active PEM control, forming a controlled Markov chain where switching is managed rather than triggered by boundary states. The population dynamics are given by
\begin{equation}
q[k+1] = M \left(I + M_{\beta_{\text{on}}[k]}^{+} - M_{\beta_{\text{off}}^{-}[k]}^{-} \right) q[k],
\label{eq:controlled model}
\end{equation}
where the matrices \( M_{\beta_{\text{on}}[k]}^{+} \) and \( M_{\beta_{\text{off}}^{-}[k]}^{-} \) encode the influence of control actions that shift devices between ON and OFF-states. \(\beta_{\text{on}}[k]\) and \(\beta_{\text{off}}^{-}[k]\) defined as the proportion of ON and OFF-requests respectively accepted by the controller and are functions of the tracking error at time instant $k$. As such both \(\beta_{\text{on}}[k]\) and \(\beta_{\text{off}}^{-}[k]\) $\in [0,1]$. The following sections detail how ~\eqref{eq:controlled model} as controlled model is able to capture PEM-based control at an aggregate level.

\subsection{Modeling turn ON requests}

Under PEM, the probability of making a turn-ON request is a function of temperature as given by ~\eqref{eq:turn-on}, as such due to discretization of temperature space, each OFF-device in the same temperature bin is equally likely to submit a turn-ON request based on the bin midpoint temperature.

With \( \beta_{\text{on}}[k] \) being the fraction of total turn-ON request accepted and $p_{\text{req}}^{i}$ denoting the turn-ON request probability for bin $i$ governed by \eqref{p_turn_on}, the expected proportion of devices turning ON from bin $i$ is given by \( \beta_{\text{on}}[k] \, p_{\text{req}}^{i} \, q_{\text{on}}^{i}[k] \).
More generally, we can define $T_{\text{req}} := \text{diag}\left\{p_{\text{req}}^{1}, \dots, p_{\text{req}}^{N}\right\} \in \mathbb{R}^{N \times N}$. Then we define \( q^{+}[k] \in \mathbb{R}^{2N} \) to denote the fraction of OFF-devices turning ON from temperature bins $q[k]$ as
\begin{equation}
q^{+}[k] := 
\begin{pmatrix} 
0_N & \beta_{\text{on}}[k] \, T_{\text{req}} \\[6pt]
0_N & -\beta_{\text{on}}[k] \, T_{\text{req}} 
\end{pmatrix} q[k] = M_{\beta_{\text{on}}[k]}^{+} q[k],
\end{equation} 
where $M_{\beta_{\text{on}}[k]}^{+} \in \mathbb{R}^{2N \times 2N}$ can be then directly used in~\eqref{eq:controlled model} to model the transition of devices from OFF-to-ON states.  

The other important signal of PEM coordination methodology is the aggregate ON-requests being sent from the OFF-devices, which is denoted $n_{\text{req}}^{\text{on}}[k]$. Specifically, the aggregate turn-ON requests at time $k$ is given by:
\begin{equation}
  n_{\text{req}}^{\text{on}}[k] = \mathbb{\textbf{1}}_{N }^{T} T_\text{req}{q}_\text{off}[k].
\end{equation}

Consequently, when a fraction of devices in the respective temperature bins is turned ON, they are simultaneously moved into a deterministic timer, consisting of \(\left\lceil \frac{\delta}{\Delta t} \right\rceil\) states.
These are referred to as the timer bins or states, where \( \delta \) is the epoch (or packet length) and \( \Delta t \) is the sampling time.
Let $x_p[k] \in \mathbb{R}^n$ denote the $n$ deterministic timer bins at time-step $k$. Then, the timer dynamics that describe the propagation of the ON population as it moves towards the completion of their packets as described by
\begin{equation}
x_p[k+1] = M_p x_p[k] + C_p q^{+}[k],
\label{timer_original}
\end{equation}
where $M_p \in \mathbb{R}^{n \times n}$ is a matrix with all entries set to 0 except for the first lower diagonal which are all set to 1. Matrix $C_p \in \mathbb{R}^{n \times 2N}$ places the newly turned on proportions in either the first timer bin or in a timer bin of value $j = \left\lfloor \frac{\delta - t_j}{\Delta t} \right\rfloor$ where $t_{j}$ is the time such that the bin population temperature remains below $T_{\text{max}}$ when starting from $t_{j}$ at the last timer bin. 
This concludes the modeling of turn-ON requests at an aggregate level.

\subsection{Modeling turn OFF requests}
Timer bins as introduced in \eqref{timer_original} represent the proportion of devices that are currently ON and consuming energy packets. \( x_p[k] \) denotes the ON population distribution across timer states at time step \( k \), where \( x_p^j[k] \) represents the fraction of ON devices in timer bin \( j \).

To model the likelihood of devices in timer bins making a turn OFF- request, we describe a CDF similar to \eqref{p_req_off} that assigns probability to submit OFF-requests as a function of timer bin index such that higher(lower) indices have more(less) probability to submit a turn-OFF request. To formalize this notion, we express the elapsed time since the start of energy packet consumption for devices in any timer bin \( r \in \{1, \ldots, n\} \) as \( t_{\text{elapsed}} = r \Delta t \).
To prevent frequent switching, the model sets the OFF-request probability to zero in the initial timer bins following a turn-ON event. For AC units, this is governed by the compressor lockout time, ensuring a minimum ON duration before turn-OFF requests are allowed to be submitted.

Let the compressor lock out time be denoted by $t_{\text{lock-out}}$ and bin index $r_{\text{lo}} \in \{1,\hdots, n\}$ be the smallest index such that $r_{\text{lo}} \Delta t \ge t_{\textit{lock-out}}$  Then the CDF for making turn-OFF requests by the proportion of devices in the bin $r$ is given by $P_{\text{req-off}}(r) = 1-e^{-\mu_{\text{off}}^{r} \Delta t}$, where
\begin{equation}
\mu_{\text{off}}^{r} = 
\begin{cases} 
0, & \text{if } t_{\text{elapsed}} \le t_{\text{lock-out}}, \\[6pt]
\frac{r - r_{\text{lo}}}{n - r} m_{\text{R-off}}, & \text{if } t_{\text{lock-out}} < t_{\text{elapsed}} < n \Delta t,
\end{cases}
\label{packet_lengths}
\end{equation}
and $m_{\text{R-off}}$ depends inversely on the $\text{MTTR}^{\text{off}}$ which is the time a device must wait before making another OFF-request after a rejection. It is worth mentioning that the entire proportion of devices in the last timer bin will turn-OFF due to the expiration of the packet. Then, the probability of each bin making an OFF-request is defined by turn-OFF probability matrix
\begin{align}
M_{\text{off}} := \text{diag}\{[\mathbf{0}_{r_\text{lo}}^\top, P_\text{req-off}^{r_\text{lo}+1},P_\text{req-off}^{r_\text{lo}+2},\hdots, P_\text{req-off}^n]\},
\end{align}
which is a time-invariant diagonal $n\times n$ matrix with entries $P_{\text{req-off}}^{j} = 1 - e^{-\mu_{\text{off}}^{j} \Delta t}$, for all $j \in \{r_{\text{lo}}+1, r_{\text{lo}}+2, \ldots, n\}$, representing the probability of making an OFF-request in timer bin $j$.
Thus, the expected fraction of the devices in timer bin $j$ making the request to turn-OFF is given by $P_{\text{req-off}}^{j} \, x_{p}^{j}[k]$, which in vector form is
\begin{equation}
\hat{x}_p^{\text{off}}[k] = M_{\text{off}} \, x_{p}[k].
\label{eq:turnoff}
\end{equation}
Since, $\beta_{\text{off}}[k]$ fraction of incoming OFF-requests that get accepted from each bin, the expected fraction of devices turning-OFF from timer bin $j$ can then be given as $\beta_{\text{off}}[k] P_{\text{req-off}}^{j} \hspace{0.1cm} x_{p}^{j}[k]$. Consequently, the same can be represented for all the timer bins as
\begin{equation}
x_{p}^{\text{off}}[k] = \beta_{\text{off}}[k] \hat{x}_p^{\text{off}}[k].
\end{equation}
where $x_{p}^{\text{off}}[k] \in \mathbb{R}^{n}$ represents proportion of devices turning-OFF across timer bins at time $k$. 
Also 
$\hat{x}_p^{\text{off}}$ as obtained in \eqref{eq:turnoff} can then be used to calculate the aggregate turn-OFF request from the population at time $k$ as
\begin{equation}
n_{\text{req}}^{\text{off}}[k] = \mathbf{1}_{n}^{T} \hat{x}_p^{\text{off}}[k].
\label{total_turn_off}
\end{equation} 
Preceding discussion implies a change in the propagation of the timer dynamics as all the population proportion at time instant $k$ in the timer bins may or may not be moving to the next subsequent bins at time $k+1$. Consequently, we update~\eqref{timer_original} as
\begin{equation}
x_{p}[k+1] = M_{p}\, x_{p}[k] + C_{p}\, q^{+}[k] - M_p x_p^\text{off}[k]
\end{equation}
to account for accepted turn-OFF requests.

In order to map the ON devices transitioning from the timer bins to OFF temperature bins, we define  the fraction of devices turning OFF with respect to the total ON population as
\begin{equation}
\beta_{\text{off}}^{-}[k] = \frac{\beta_\text{off}[k] \mathbb{\textbf{1}}_{n}^{T} M_\text{off} {x}_{p}[k]}{\mathbb{\textbf{1}}_{n}^{T} x_{p}[k]} + \frac{(1 - \beta_{\text{off}}[k]) x_{p}^{n}[k]}{{\mathbb{\textbf{1}}_{n}^{T} x_{p}[k]}},
\label{beta eq}
\end{equation}
which enables the mapping back from timer states to temperature states in~\eqref{eq:controlled model}. The first term in~\eqref{beta eq} represents the OFF-to-ON transitions from accepted OFF requests, while the second term accounts for completed packets in the last timer bin whose requests were not accepted. Specifically, since all devices in the final timer bin are guaranteed to turn OFF, any value of $\beta_{\text{off}}[k] \in [0,1)$ implies that a fraction $(1 - \beta_{\text{off}}[k])$ of the final timer bin would otherwise be incorrectly retained in the ON state. Hence instead of directly using the value of $\beta_{\text{off}}[k]$ in \eqref{eq:controlled model} we calculate an effective value of $\beta^{-}_{\text{off}}[k]$ which is used with~\eqref{eq:controlled model}

Since \(\beta_{\text{off}}^{-}[k]\) is defined in the timer domain and the exact distribution of devices in a timer bin across temperature bins at time step \(k\) is unknown, it is assumed that the averaging effect distributes the OFF transitions uniformly across temperature bins when converting back from timer to temperature space. The value of \(\beta_{\text{off}}^{-}[k]\) can then be mapped to temperature space as
\begin{equation}
q^{-}[k] = 
\begin{pmatrix} 
\beta_{\text{off}}^{-}[k] I_{N} & 0_N \\[6pt]
-\beta_{\text{off}}^{-}[k] I_{N} & 0_N 
\end{pmatrix} q[k] = M_{\beta_{\text{off}}^{-}[k]}^{-} q[k],
\end{equation}
which can be then used directly in \eqref{eq:controlled model} and thus completes the modeling of the turn OFF-requests.

\subsection{Modeling opt-out mode of PEM}
The opt out mode of PEM is incorporated to account for devices that may temporarily operate outside the deadband region. To maintain the accuracy of the model, it is essential to carefully design the number of opt-out states based on device parameters and sampling time. Insufficient representation of these states could lead to modeling errors. Inclusion of the opt-out dynamics requires a very simple augmentation of \eqref{eq:controlled model} by including a new matrix $M_{\text{exit}}$ which is further made up of sub matrices. The equations governing the population dynamics then become
\begin{equation}
q[k+1] = M_{\text{exit}} \, \bar{M} \, q[k],
\label{complete model}
\end{equation}
where $q[k]=(q_{\text{on}}^{\text{opt out}}[k] \hspace{0.1cm} q_{\text{off}}^{\text{opt out}}[k] \hspace{0.1cm} q_{\text{on}}[k] \hspace{0.1cm} q_{\text{off}}[k])^\top$ is now augmented to include the opt-outs.
Also, the matrices $\bar{M}$ and $M_{\text{exit}}$ are given as 
\begin{equation}
\bar{M} = 
\begin{pmatrix} 
I_z & 0 \\[6pt]
0_N & I + M_{\beta_{\text{on}}[k]}^{+} - M_{\beta_{\text{off}}[k]}^{-}
\end{pmatrix}. 
\end{equation}
As devices opt-out in PEM they remain locked as either ON/OFF and hence unavailable to participate in the tracking till the temperature specified for re-entry in the PEM is achieved by the devices. This effectively makes them unavailable for PEM coordinator (since they will not provide any requests), which has a direct implication on the fleet's ability to regulate the aggregate power and, hence, its tracking capability. The specified temperature bin for which opted-out devices return to PEM scheme is a design parameter in the aggregate model. This choice also determines the size of the vector \((q_{\text{on}}^{\text{opt out}}[k] \hspace{0.1cm} q_{\text{off}}^{\text{opt out}}[k])^\top\), as well as the matrices \(I_{z}\) and \(M_{\text{exit}}\) in the aggregate model. The matrix \(M_{\text{exit}}\), as introduced in \eqref{complete model}, controls the transitions of the population proportion between opt-out states and their entry and exit from PEM. For more details on opt-out dynamics, please see~\cite{main_macro}.   
\subsection{Aggregate power control policy}
The effectiveness of any DER coordination method can be characterized by reference signal tracking accuracy, A DER coordinator must manage the competing objectives of minimizing tracking error while maintaining customer comfort. In PEM, this trade-off is governed by control parameters $\beta_{\text{on}}$ and $\beta_{\text{off}}$, computed from the tracking error. While any incremental or decremental change in the reference power signal can be tracked by accepting appropriate combination of ON and OFF requests, this approach can increase the switching frequency of devices. As such an optimization problem could be set up to determine the values of $\beta_{\text{on}}$ and $\beta_{\text{off}}$ that maximize comfort. However here, a complementarity constraint $\beta_{\text{on}}[k]\,\beta_{\text{off}}[k]=0$ ensures only one request type is accepted per time step, minimizing switching. As a result, an increase in the reference signal is handled by accepting ON-requests, while a decrease is handled via OFF-requests.

The aggregate power demand of the fleet $P_{\text{agg}}[k]$ at any time $k$ can be given as 
\begin{equation}
    P_{\text{agg}}[k] = P_{\text{rate}} C^{\top}q[k],
\end{equation}
where 
\( C = \left( \mathbf{1}_{\text{opt-on}} \hspace{0.2cm} \mathbf{0}_{\text{opt-off}} \hspace{0.2cm} \mathbf{1}_{N} \hspace{0.2cm} \mathbf{0}_{N} \right)^{\top} \),  
and the dimensions of \( \mathbf{1}_{\text{opt-on}} \) and \( \mathbf{0}_{\text{opt-off}} \) depend on the return states chosen for the opt-out devices.

The aggregate power of the fleet evolves as 
\begin{align}
    P_{\text{agg}}[k+1] =\ & P_{\text{agg}}[k] + \left( \beta_{\text{on}}[k]\,n_{\text{req}}^{\text{on}}[k] - \beta_{\text{off}}\,n_{\text{req}}^{\text{off}}[k] \right) P_{\text{rate}} \nonumber \\
    \quad & - (1 - \beta_{\text{off}}[k])\,x_{p}^{n}[k]\,P_{\text{rate}}.
    \label{aggregate power}
\end{align}
Minimization of the tracking error \( P_{\text{ref}}[k+1] - P_{\text{agg}}[k+1] \) determines the number of ON and OFF requests that must be accepted, giving rise to two scenarios.
If \( P_{\text{ref}}[k+1] > P_{\text{agg}}[k] \), the PEM coordinator needs to increase fleet power consumption at time $k$. Under the complementarity constraint, we set \( \beta_{\text{off}}[k] = 0 \), implying that no OFF requests are accepted at time $k$. Then, equation~\eqref{aggregate power} reduces to:
\begin{equation}
    P_{\text{agg}}[k+1] = P_{\text{agg}}[k] + \beta_{\text{on}}[k]\, n_{\text{req}}^{\text{on}}[k]\, P_{\text{rate}} - x_{p}^{n}[k]\, P_{\text{rate}}.
    \label{aggregate_power_on_case}
\end{equation}

Accordingly, the value of $\beta_{\text{on}}[k]$ to meet the desired demand increase is given by:
\begin{equation}
\beta_{\text{on}}[k] = \min\left\{ 1,\ \frac{P_{\text{ref}}[k+1] - P_{\text{agg}}[k] + x_p^n[k]\,P_{\text{rate}}}{P_{\text{rate}}\,n_{\text{req}}^{\text{on}}[k]} \right\}.
\end{equation}
Similarly, if \( P_{\text{agg}}[k] > P_{\text{ref}}[k+1] \), the PEM coordinator must reduce the fleet's power consumption. In this case, we set \( \beta_{\text{on}}[k] = 0 \), meaning no ON requests are accepted at time $k$. Then, equation~\eqref{aggregate power} reduces to:
\begin{align}
    P_{\text{agg}}[k+1] =\ & P_{\text{agg}}[k] - \beta_{\text{off}}[k]\,n_{\text{req}}^{\text{off}}[k]\,P_{\text{rate}} \nonumber \\
    & - (1 - \beta_{\text{off}}[k])\,x_{p}^{n}[k]\,P_{\text{rate}}.
    \label{aggregate_power_off_case}
\end{align}

Solving for the $\beta_{\text{off}}[k]$ to determine the accepted OFF- proportion of requests to meet the desired demand reduction:
\begin{equation}
\beta_{\text{off}}[k] = \min\left\{ 1,\ \frac{P_{\text{agg}}[k] - P_{\text{ref}}[k+1] - x_p^n[k]\,P_{\text{rate}}}{P_{\text{rate}} \left( n_{\text{req}}^{\text{off}}[k] - x_p^n[k] \right)} \right\}.
\label{betaoff}
\end{equation}
This completes the development of the \textit{enhanced aggregate model} that captures complementary ON/OFF request from the fleet. In the next section, we compare the macro and micro models to validate  accuracy and performance for different test cases. 
\begin{figure*}[t]
    \centering

    \begin{subfigure}{0.4\textwidth}  
        \includegraphics[width=\linewidth]{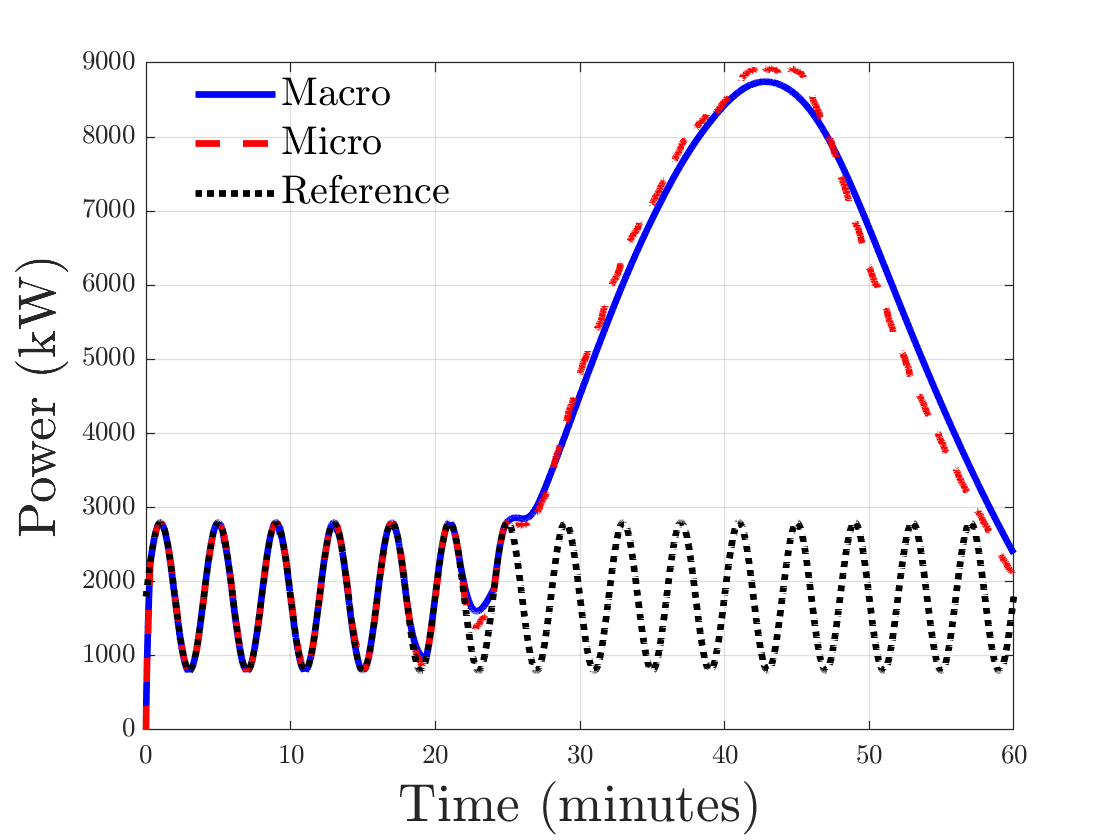}
        \caption{Aggregate power during reference tracking.}
    \end{subfigure}
    \hfill
    \begin{subfigure}{0.4\textwidth}
        \includegraphics[width=\linewidth]{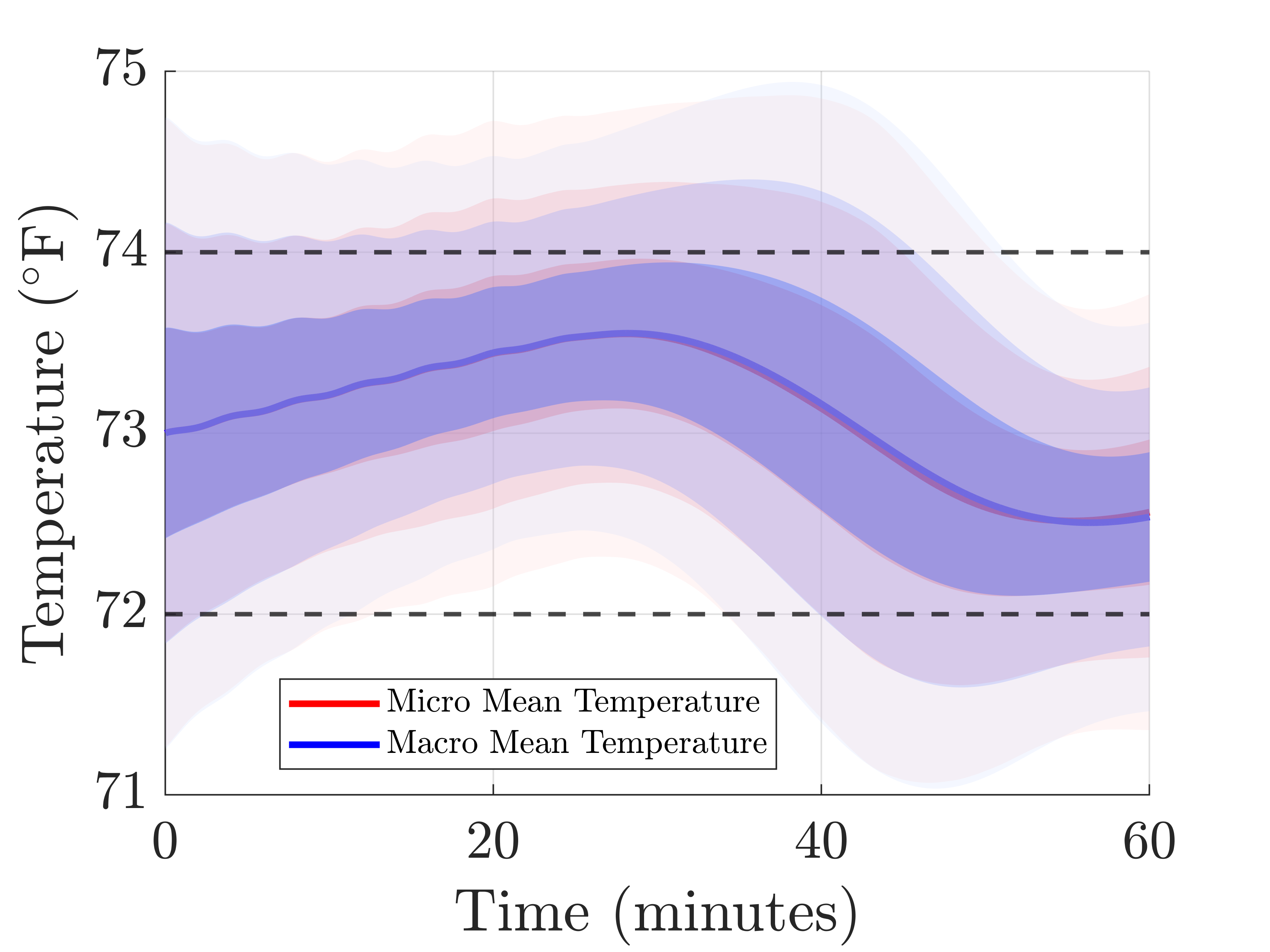}
        \caption{Population Temperature Statistics}
    \end{subfigure}

    \vspace{0.6em}  

    \begin{subfigure}{0.4\textwidth}
        \includegraphics[width=\linewidth]{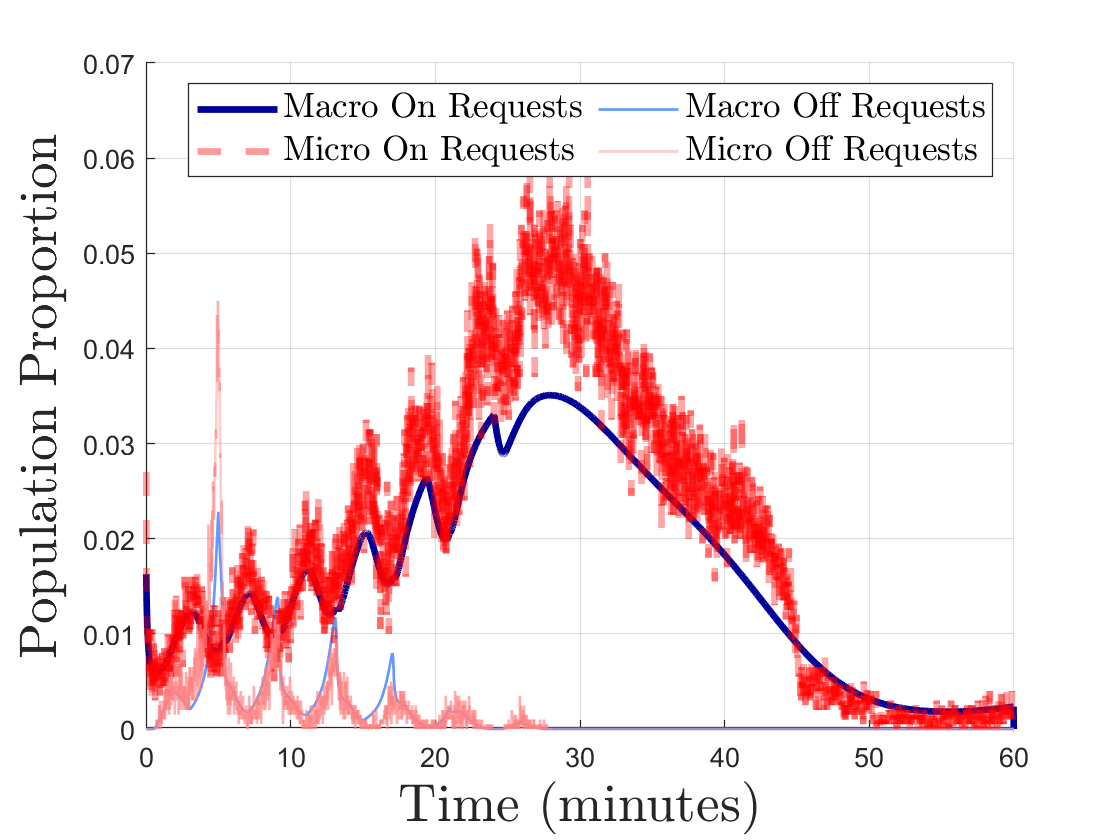}
        \caption{Proportion of devices requesting ON and OFF packets.}
    \end{subfigure}
    \hfill
    \begin{subfigure}{0.4\textwidth}
        \includegraphics[width=\linewidth]{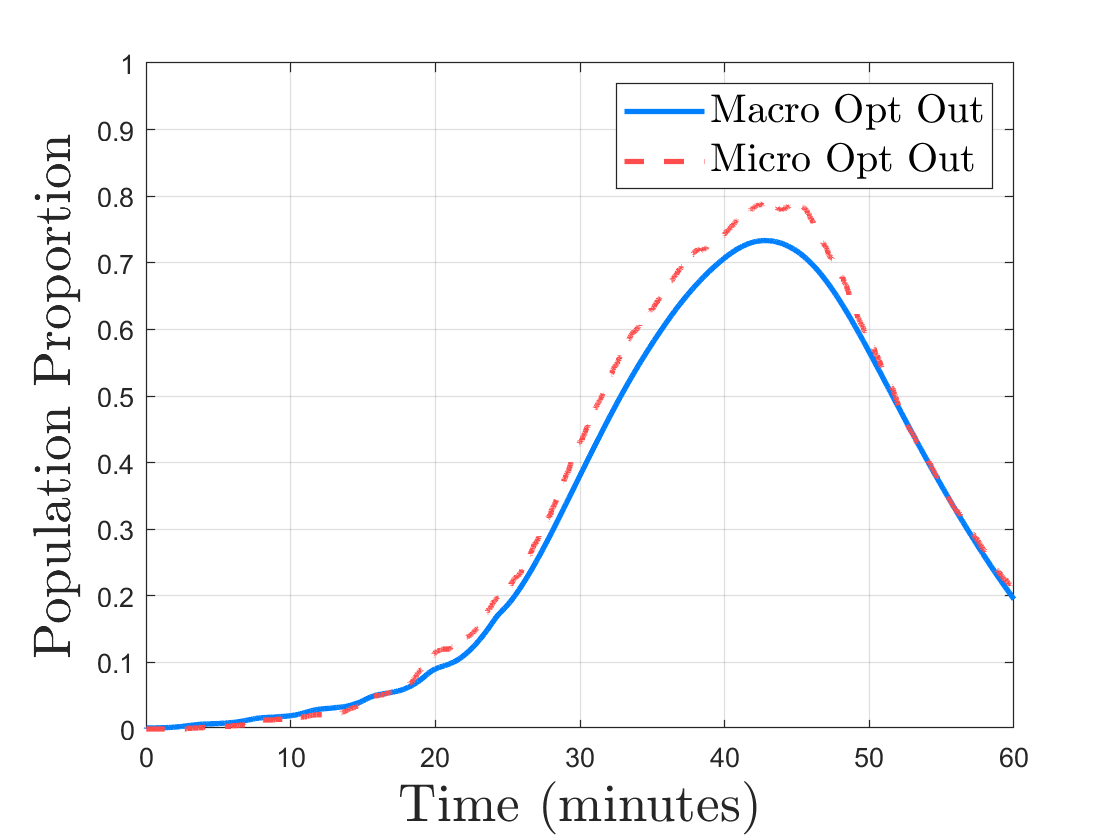}
        \caption{Proportion of devices opting Out of PEM.}
    \end{subfigure}
    \vspace{6pt}
    \caption{Model validation showing power trajectory, temperature distribution, request patterns, and opt-out dynamics over one hour for a fleet of 2000 AC units.}
    \label{fig:model-validation}
\end{figure*}
   
\section{Simulation-based Validation \& Test-cases}
\noindent In this section, 
simulation-based analysis highlights model accuracy across states, request patterns, opt-out dynamics and aggregate power reference tracking capability. The enhanced macro model is compared against an agent-based (micro) model/simulation involving AC units to determine model accuracy. All simulations and test cases use AC unit parameters derived from~\cite{leke} with an ambient temperature at $89^{\circ}\mathrm{F}$ ,a setpoint of $73^{\circ}\mathrm{F}$ and rated power of 6~kW unless stated otherwise. The parameters of thermal resistance and capacitance are modeled with variation of $\pm 5\%$ sampled from a uniform distribution to capture heterogeneity of a group of similar residential homes and AC units. Additionally, we show how the enhanced macro model's ability to capture off-requests (which changes the effective packet lengths) can be used to  inform design of randomized packet lengths for conventional PEM~\cite{braham}.


\subsection{Enhanced macro model validation}\label{sec_macroVal}
\begin{figure}
    \centering
    \includegraphics[width=1\linewidth]{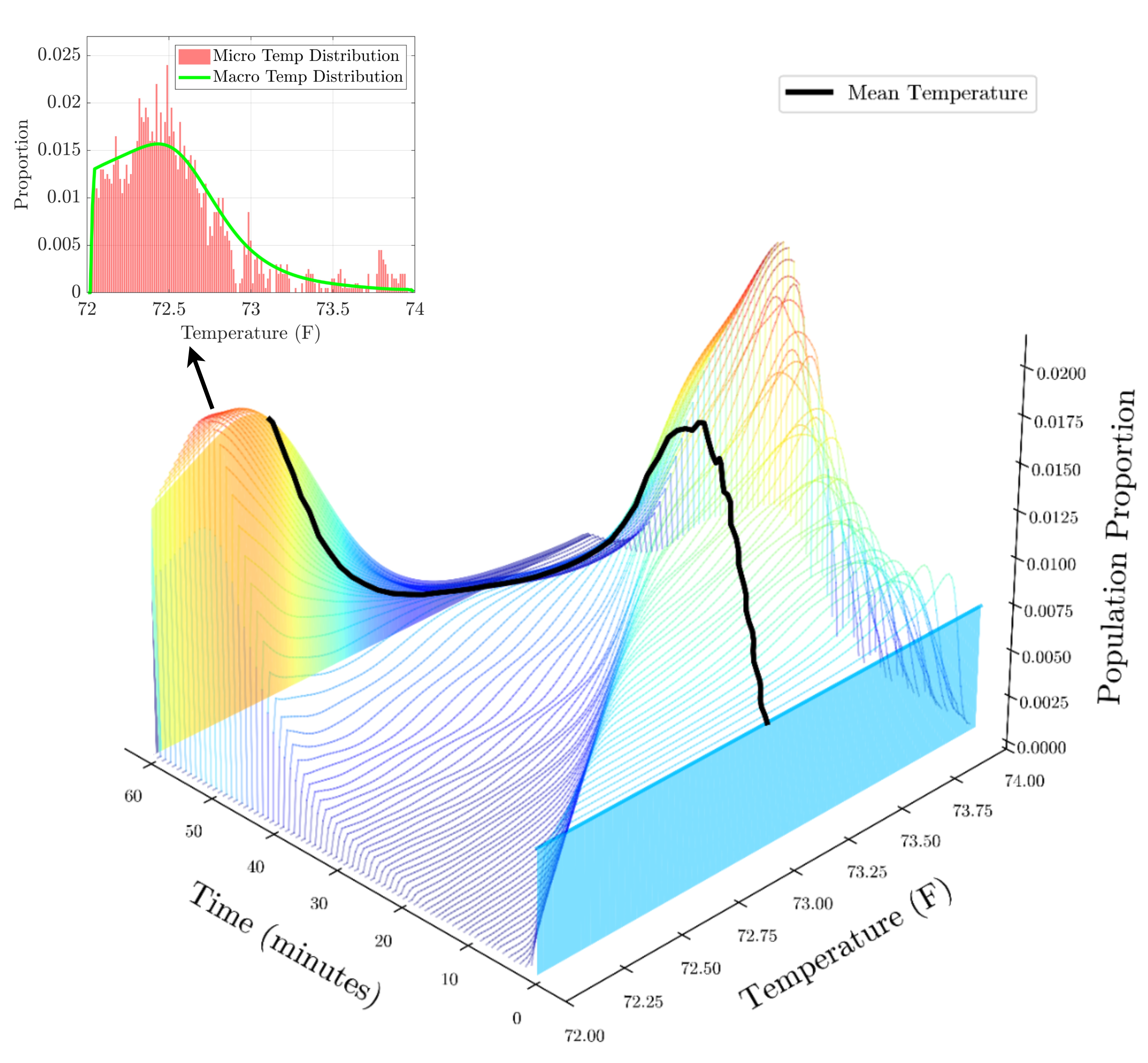}
    \caption{Temporal evolution of macro model temperature distributions and their mean trajectory}
    \label{fig:temporal evolution}
\end{figure}

We simulate a population of 2000 ACs over one hour, initialized as OFF with uniformly distributed temperatures across state bins. A sinusoidal signal 
$P_{\text{ref}}(t):= 1000~\sin\left( \frac{\pi}{120} \, t \right) + 1800$~kW
is employed as a reference signal. 
The simulations presented in Fig.~\ref{fig:model-validation} compare the enhanced macro model against its corresponding agent-based simulation for states, outputs, and other characteristics of PEM. A reference signal with a mean below the fleet’s nominal power consumption was deliberately selected to evaluate model performance under edge-case control scenarios. Initially, the fleet has sufficient energy due to a uniform temperature distribution and the mean temperature of the fleet at setpoint, allowing effective manipulation of ON and OFF-requests to accurately track the reference signal. However, as time progresses, the mean temperature drifts toward the upper limit of the deadband, as illustrated in Fig.~\ref{fig:temporal evolution}, indicating a reduced ability of the fleet to adjust aggregate power and a gradual loss of its effective state of charge.
As observed in Fig.~\ref{fig:model-validation} at the 22-minute mark, a significant portion of devices opt out as the temperatures rise due to repeated denial of turn ON requests, impairing tracking performance.The fleet cannot track the reference signal until opted-out devices rejoin the PEM scheme.

Importantly, the enhanced macro model can accurately predict the point at which the micro model (fleet) no longer tracks the reference signal. The RMSE between the macro and micro models' power trajectories is 198~kW. The micro and macro temperature populations are compared using the 2-norm and $\max$-norm of the mean temperature differences and the differences of temperature standard deviations over all time steps\footnote{With slight abuse of notation, the metrics are defined as $||\hat{\mu}_\text{macro} - \hat{\mu}_\text{micro}||_p$ and $||\hat{\sigma}_\text{macro} - \hat{\sigma}_\text{micro}||_p$ for $p\in \{2,\infty\}$, where mean vectors $\hat{\mu} :=[\mu[1],\hdots, \mu[K]]^\top$ and standard deviation vectors $\hat{\sigma} :=[\sigma[1],\hdots, \sigma[K]]^\top$ are given for  micro and macro model populations.}. Results for the macro model are summarized in  Table~\ref{state accuracy metrics} and demonstrate accurate estimates of population statistics. Similarly, we observed agreement in both the aggregate request patterns and the fraction of devices opting out over time; these two metrics are critical for the aggregate PEM model to accurately capture and predict fleet performance. Typical modeling errors arise from temperature discretization, fleet heterogeneity, and rounding off when translating micro-model statistics to a bin-based representation for comparison with the macro model.
\begin{table}[h]
\centering
\caption{Temperature accuracy metrics in $^\circ F$ }
\renewcommand{\arraystretch}{1.2}
\begin{tabularx}{0.8\columnwidth}{l *{2}{>{\centering\arraybackslash}X}}
\toprule
 & \textbf{Mean Temperature} & \textbf{Standard Deviation} \\
\midrule
2-norm & 0.0511 & 0.0362 \\
$\infty$-norm & 0.0870 & 0.0753 \\
\bottomrule
\end{tabularx}
\label{state accuracy metrics}
\end{table}
Fig.~\ref{fig:temporal evolution} illustrates the temporal evolution of temperature distributions in the enhanced macro model, highlighting their correlation with the aggregate power trajectory. Post 22 minute mark we see the mean population temperature starting to shift back towards set point of 73$^{\circ}\mathrm{F}$ as higher proportion of devices move towards the colder temperature region as they traverse through the ON opt-out mode reflected by the increase in the aggregate demand. The figure also shows the final time step temperature distributions of the macro and agent-based models, with good visual agreement and a Pearson correlation of 0.9 further showcasing model accuracy. At the final time step, the macro model recorded a mean temperature of 72.5$^{\circ}\mathrm{F}$ and a standard deviation of 0.36$^{\circ}\mathrm{F}$, closely matching the agent-based model’s mean of 72.6$^{\circ}\mathrm{F}$ and standard deviation of 0.40$^{\circ}\mathrm{F}$, reinforcing the strong agreement between the simulations.

\subsection{Reference tracking with a Reg-D signal}
The enhanced macro model was evaluated alongside its equivalent agent-based simulation using ten distinct PJM Reg-D AGC signals as reference, each spanning one hour and sampled at 2-second intervals. These signals were randomly selected from historical data covering a six-month period. The tracking performance was assessed using a fleet of 1000 AC units. The initial probability mass vector $q[0]$ was set to the fleet’s steady-state distribution, derived either as the eigenvector of the transition matrix (eigenvalue 1) or from simulation. The AGC signals, sourced from \cite{pjm_ancillary_services}, were scaled to 1,800 kW  approximating the fleet's nominal power consumption. Fig. 5 illustrates the tracking performance for one representative signal, alongside corresponding micro model and their mean temperatures over time. 

\begin{figure}[t]
    \centering
    \begin{subfigure}{1\linewidth}
        \includegraphics[width=1\linewidth]{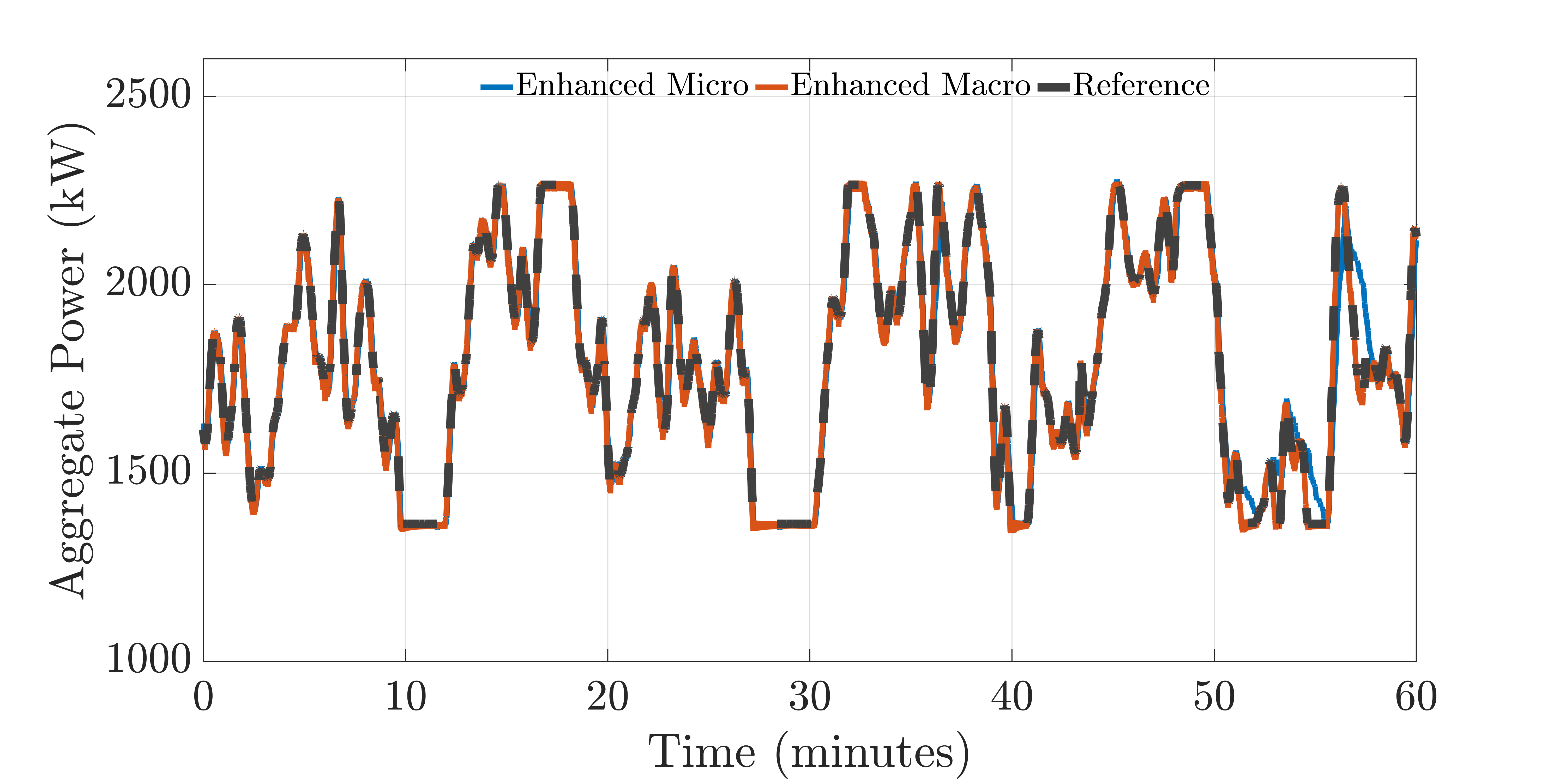}
        \label{fig:power-tracking}
    \end{subfigure}
    \begin{subfigure}{1\linewidth}
        \includegraphics[width=1\linewidth]{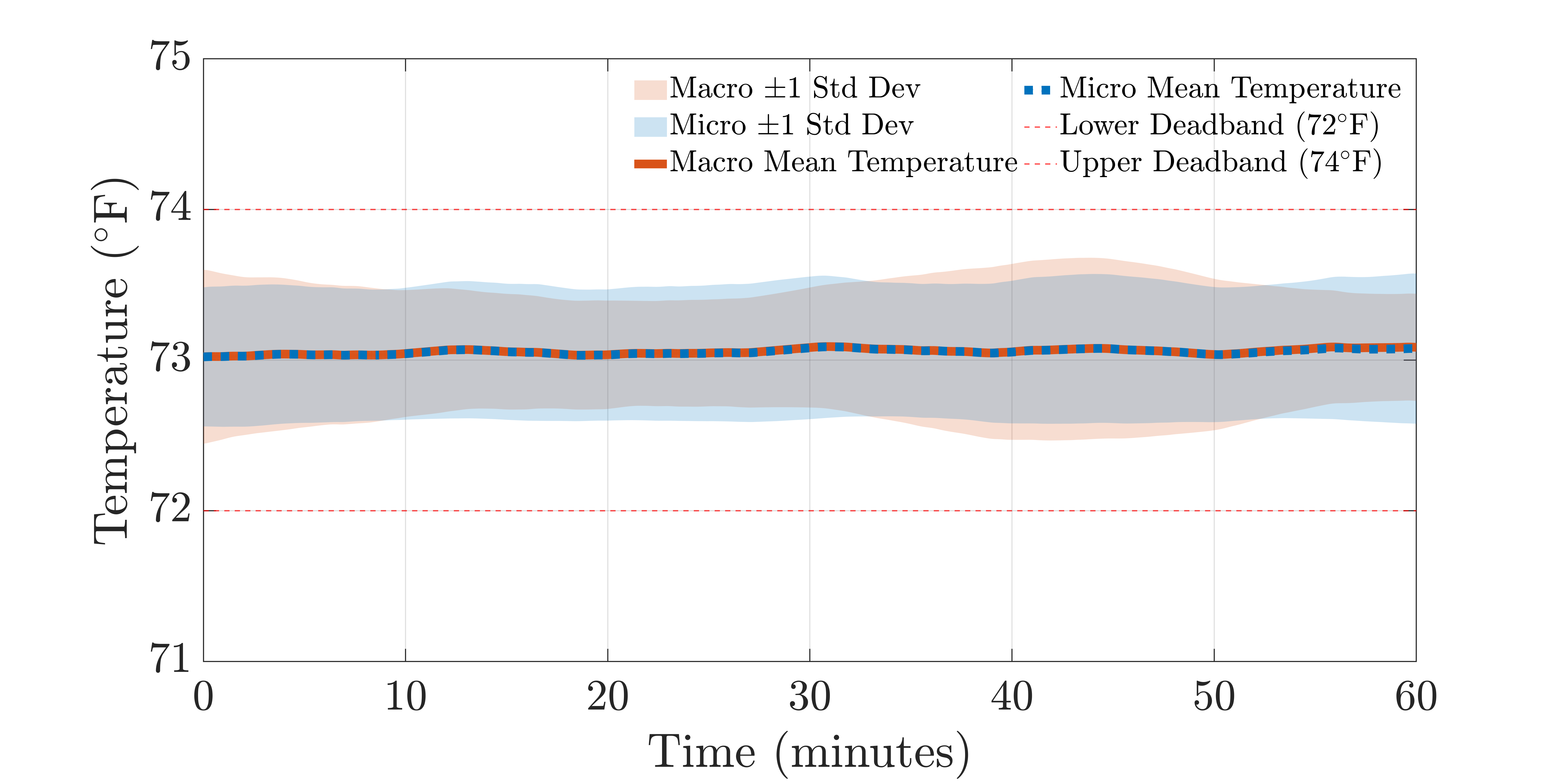}
        \label{fig:temperature-tracking}
    \end{subfigure}
    \caption{Performance of the enhanced macro model while tracking an AGC Reg-D signal. Top: power tracking comparison.   Bottom: the average population temperatures during Reg-D tracking.}
    \label{fig:tracking-performance}
\end{figure}

From Fig.~\ref{fig:tracking-performance}, the first notable observation is the close agreement between the enhanced macro model and its corresponding agent based simulation while tracking Reg D signal. This agreement is quantified using the RMSE between the two signals for both mean temperature and aggregate power trajectories. The errors were reported to be 0.04$^{\circ}\mathrm{F}$ and 57.14 kW, respectively. 
Thus, the enhanced macro model is able to retain the model accuracy reasonably well when tested with a rather random Reg-D signal. Having established accurate fleet-state predictions, the enhanced macro model can further estimate the fleet’s tracking capability. To obtain this accuracy estimate, we compare the RMSE of aggregate power of both enhanced macro model and its agent based simulation with reference signal. For this simulation, the enhanced macro model predicts  tracking with an RMSE of 20.67 kW, while the actual accuracy achieved by the fleet as shown by the micro model was with an error of 50.85 kW. This suggests that, despite modeling simplifications and assumptions, the macro model offers a reasonably accurate prediction about the tracking capabilities of the TCL fleet.
\subsection{Macro model robustness analysis} The macro model uses average parameters to represent the heterogeneous population in a single Markov model, which introduces errors relative to agent-level behavior of the micro model. It is expected that the greater the parameter variance, the larger will be the discrepancy between macro and micro model simulations. While multiple Markov chains can reduce the effects of heterogeneous devices by grouping devices with similar parameters \cite{NazirHiskens17_NoiseParameterHeterogeneityInAggregateModelsOfThermostatically}, our macro model uses a single Markov chain. We would like to examine the robustness of this macro model in accommodating parameter heterogeneity. Specifically, the micro model samples $R^{n}_{\text{eq}}$, $C^{n}_{\text{eq}}$, and $P^{n}_{\text{rate}}$ from uniform distributions with increasing variances and quantify the model deviations through the Kullback-Leibler divergence (KLD) of temperature distributions across all time steps for each simulations. That is, for each increase in variance, we compute the mean KLD over $K$ time steps of simulation as
\vspace{-1 em}  
\begin{equation}
\overline{KL}(q_{\text{macro}}\|q_{\text{micro}}) :=
\frac{1}{K} \sum_{k=1}^{K} \sum_{i=1}^{N_b} q_{\text{macro},i}[k]
\log \frac{q_{\text{macro},i}[k]}{q_{\text{micro},i}[k]}
\end{equation}
\begin{figure}
    \centering
    \includegraphics[width=1\linewidth]{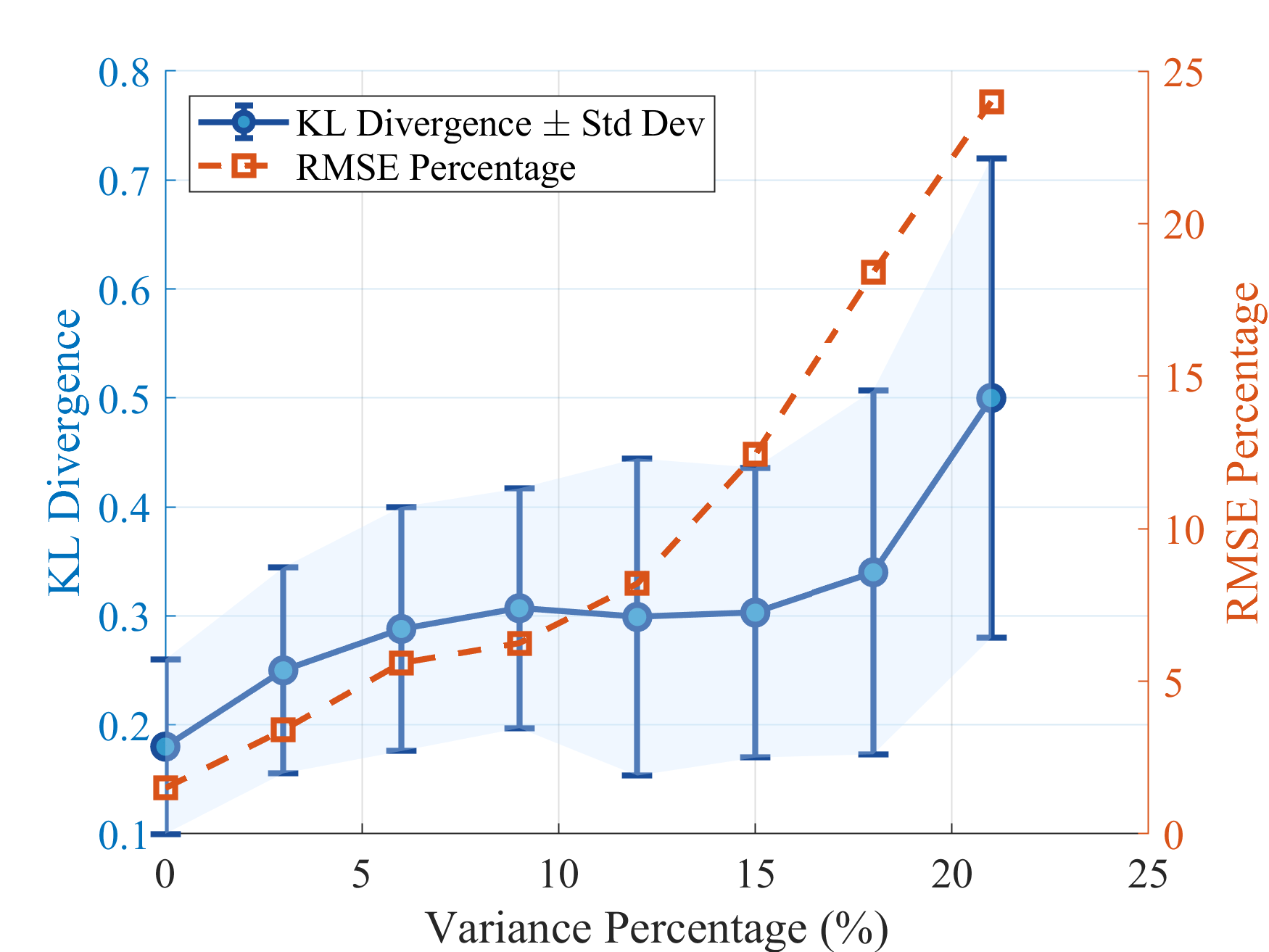}
    \caption{Variation of KL divergence  and RMSE of power outputs between macro and micro models with increasing variance of the parameters}
    \label{kl divergence}
\end{figure}
where $q_{\text{macro},i}[k]$ and $q_{\text{micro},i}[k]$ are temperature bin $i$ at time $k$ of the macro and micro models, respectively. Fig.~\ref{kl divergence} shows the mean and $\pm~1$ standard deviation of the KL divergence measure over the 1-hour simulation interval. We run multiple simulations per variance point and use the worst case values in the plot.  It is seen that increasing the parameter variance, generally increases KL metrics. However, Fig.~\ref{kl divergence} has a rather flat region between 8-15\%, which can be attributed i) using only a one-hour simulation window; and ii) a limited number of parameter samples. Specifically, the increase in variance does not imply that all parameters are sampled near their extremes. Note that in the case when the parameter variance is zero (i.e., homogeneous fleet), $\overline {KL} > 0$. This is due to the discretization errors inherent to the macro model. 
Finally, the increasing KLD with higher parameter variance aligns with larger power mismatches between micro and macro models, measured by \% RMSE. Using KLD properties, one can derive an upper limit on these mismatches, informing the acceptable level of parameter heterogeneity for robustness. For the presented model, 12\% parameter heterogeneity, corresponding to roughly 10\% power RMSE, may serve as a cutoff to claim robustness, depending on the situation.



\subsection{Variable packet lengths}
In the conventional PEM framework, each ON event is a \textit{fixed} packet epoch (e.g., 300 seconds).
Introducing turn-OFF requests allows devices to interrupt packets early, governed by $\beta_{\text{off}}$  which in  turn is based on tracking error and ramp-down needs. This results in variable packet lengths.

\begin{figure}[t]
    \centering
    \includegraphics[width=0.9\linewidth]{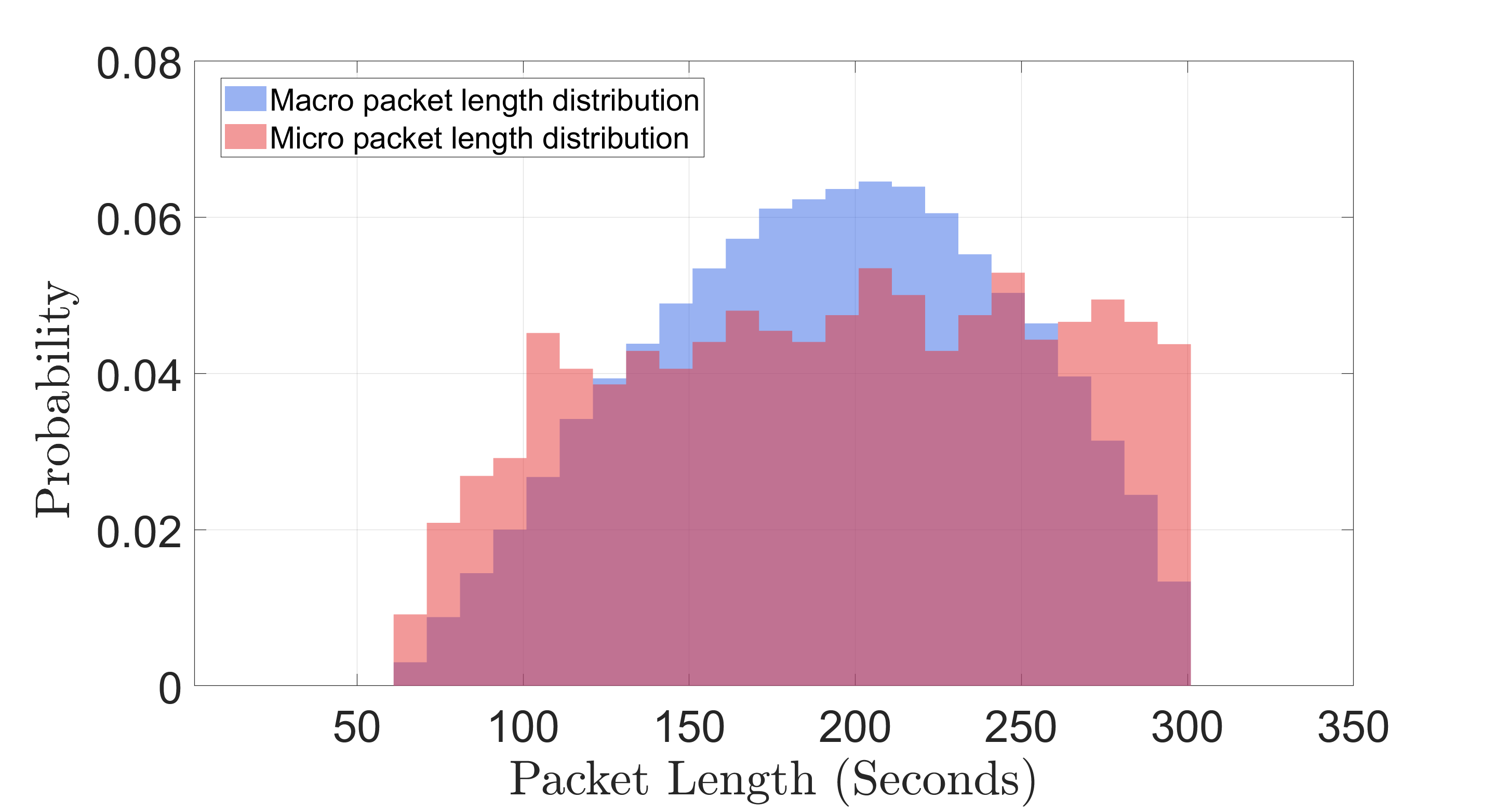}
        \caption{Packet length distributions obtained from enhanced macro and micro models, while tracking the Reg-D. Difference in the distributions is again attributed to model discretizations primarily. }
    \label{fig:distributions}
\end{figure}

Given a reference signal, one can readily obtain the distribution of packet lengths from the macro model. This distribution can then be validated against the corresponding micro model simulation, as shown in~\ref{fig:distributions} for the same Reg-D signal. 
The mean and standard deviation of the variable packet lengths for the enhanced macro/micro models were 190/194 and 53/62 seconds, respectively. These statistics further validate the accuracy of enhanced macro model. Note that packet lengths shorter than 60 s are neglected due to  device lockout, while the maximum packet length is limited to 300 s. 

The concept of variable packet lengths as implied by enhanced macro model has previously been explored in conventional PEM using uniformly distributed packet lengths for accepted turn-ON requests~\cite{braham}. However, there was no concrete basis for choosing a distribution for packet lengths.
This approach to variable (random) packet lengths was shown to improve tracking performance of the fleet relative to conventional PEM with fixed packet length.

Thus, here we are interested in   evaluating how conventional PEM with randomized packet lengths performs when informed by variable packet lengths sampled from the distribution obtained from the enhanced macro model, e.g., please see Fig.~\ref{fig:distributions}.
To evaluate the impact of variable packet lengths, we performed simulations tracking the Reg-D signal using the conventional agent-based PEM model without OFF requests (i.e., $\beta_\text{off}[k] = 0,\ \forall k$) under different packet length approaches.
Specifically, we first set up a simulation with a fixed packet length. Rather than arbitrarily selecting a value (e.g., 300~s), we use 190~s, which corresponds to the mean of the packet length distribution obtained from the enhanced macro model, to ensure a fair comparison. In the second simulation, packet lengths are instead sampled from this distribution.
Finally, we also include the tracking performance achieved by the enhanced macro model for reference.
\begin{figure}[t]
    \centering
    \includegraphics[width=0.9\linewidth]{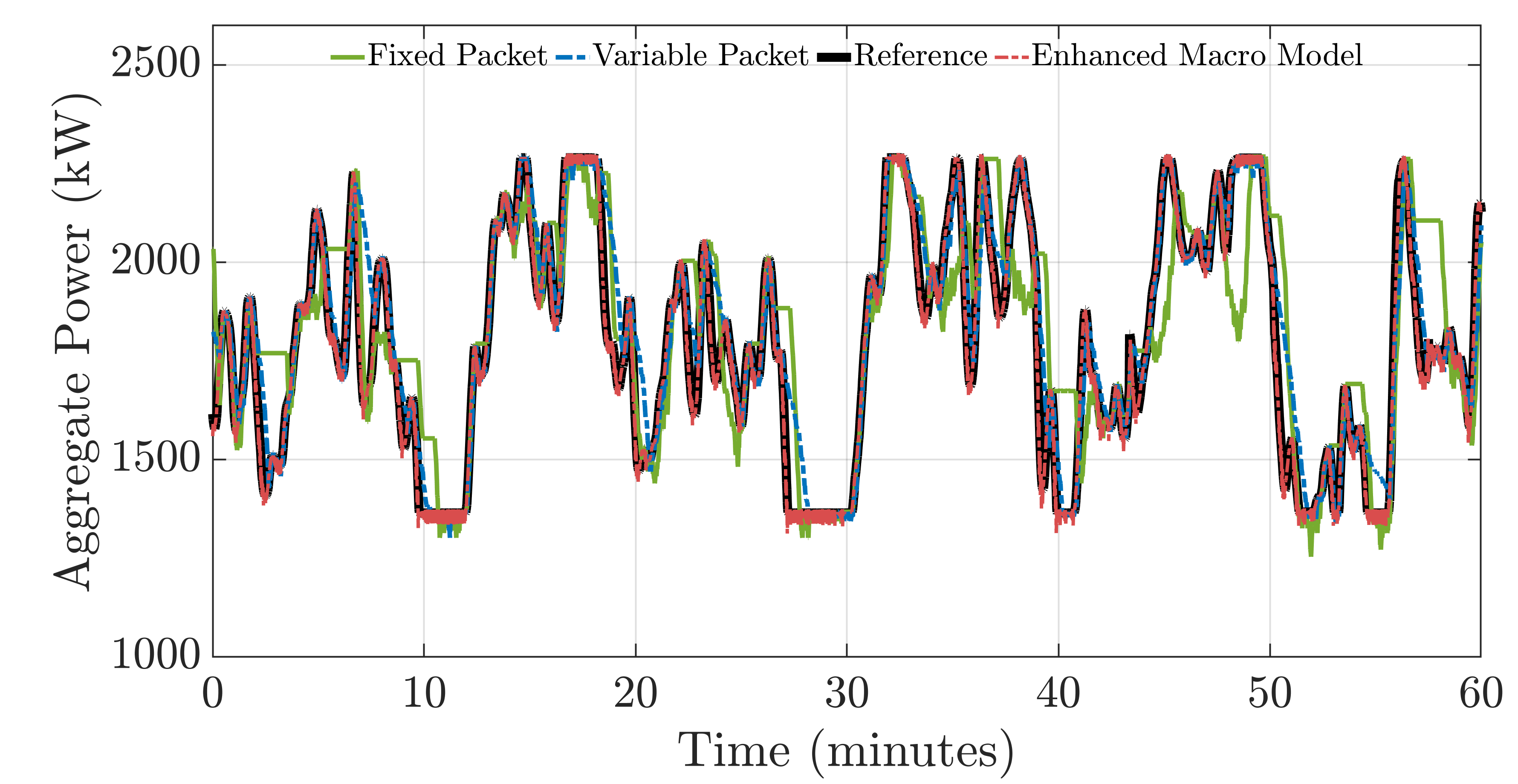}
        \caption{Improved tracking performance of conventional PEM using packet lengths sampled from distributions obtained from the enhanced macro model}
    \label{fig:improvedtracking}
\end{figure}
These reference tracking simulation results are illustrated in Fig.~\ref{fig:improvedtracking}, tracking errors are compared in Table ~\ref{tab:rmse_comparison}. As expected, the enhanced macro model outperforms both the randomized packet length model and the fixed packet length simulation for conventional PEM. Additionally, the randomized packet length simulation of conventional PEM demonstrates better tracking capability than the fixed packet PEM as expected from~\cite{braham}. This improvement is attributed to the greater availability of devices resulting from the shorter-than-average packet lengths. Higher device availability enables the population to respond more effectively to (ramp-down) changes in the reference signal, thereby enhancing tracking performance.


\begin{table}
\centering
\small
\caption{Tracking performance with Reg-D reference signal}
\renewcommand{\arraystretch}{1.3}
\begin{tabularx}{\columnwidth}{*{4}{>{\centering\arraybackslash}X}}
\toprule
\makecell{\textbf{Method}} &
\makecell{\textbf{Conv. PEM} \\ \textbf{(Fixed)}} &
\makecell{\textbf{Conv. PEM} \\ \textbf{(Random)}} &
\makecell{\textbf{Enh. PEM} \\ \textbf{(Variable)}} \\
\midrule
\mbox{\textbf{RMSE (kW)}} & 174 & 97.8 & 20.7 \\
\bottomrule
\end{tabularx}
\label{tab:rmse_comparison}
\end{table}

\section{Conclusion}
\vspace{-0.5em}
\noindent 
This paper presents a new state bin-based model that characterizes an enhanced version of PEM with both turn-on and turn-OFF requests. The proposed model, validated through different test cases, accurately predicts the population temperature dynamics, aggregate power output and key PEM characteristics. That is, the macro model provides an accurate estimate of the fleet's tracking capabilities, including edge cases, such as scenarios in which a DER fleet runs out of flexibility to track a reference signal.
We evaluate the enhanced macro model robustness under varying parameter heterogeneity using KLD metrics. Additionally, using the proposed enhanced model, we further examine variable packet lengths through a simulation-informed approach. 


Future directions will study system properties of the enhanced macro model presented in this paper, including analyzing stationarity of the population dynamics and the fleet's resulting nominal (steady-state) response. 

\printbibliography
\end{document}